\documentclass[twocolumn,aps,floatfix,show pacs]{revtex4}
\usepackage{amsmath}
\usepackage{graphicx}
\usepackage{psfrag}
\usepackage{color}
\usepackage[T1]{fontenc}
\newcommand{\ket}[1]{\left|#1\right\rangle} 
 \newcommand{\bra}[1]{\left\langle#1\right|} 

\begin{document}

\title{Few interacting fermions in one-dimensional harmonic trap}
\author{Tomasz Sowi\'nski$^{1,2}$, Tobias Grass$^{2}$, Omjyoti Dutta$^{2,3}$, and Maciej Lewenstein$^{2,4}$}
\affiliation{
\mbox{$^1$Institute of Physics of the Polish Academy of Sciences, Al. Lotnik\'ow 32/46, 02-668 Warszawa, Poland}\\
\mbox{$^2$ICFO - Institut de Ci\`ences Fot\`oniques, Parc Mediterrani de la Tecnologia, E-08860
Castelldefels, Barcelona, Spain} \\
\mbox{$^3$ Instytut Fizyki im. Mariana Smoluchowskiego, Uniwersytet Jagiello\'nski, ul. Reymonta 4, PL-30-059 Krak\'ow, Poland} \\
\mbox{$^4$ ICREA - Instituci\'o Catalana de Recerca i Estudis Avan\c cats, 08010 Barcelona, Spain}
}
\date{\today}

\begin{abstract}
We study spin-1/2 fermions, interacting via a two-body contact potential, in a
one-dimensional harmonic trap. Applying exact diagonalization, we investigate
their behavior at finite interaction strength, and discuss the role of the
ground-state degeneracy which occurs for sufficiently strong repulsive interaction.
Even low temperature or a completely depolarizing channel may then dramatically
influence the system's behavior. We calculate level occupation numbers as
signatures of thermalization, and we discuss the mechanisms to break the
degeneracy.
\end{abstract}
\maketitle

\section{Introduction}
Advances in the control and manipulation of ultracold quantum gases have opened
up a new avenue to the study of interacting few particle systems~\cite{mlbook}.
In particular, existing trapping techniques allow for exploring the physics in
low dimensions where the quantum-statistical distinction between fermionic and
bosonic particles experiences severe modifications. A striking property
of one-dimensional systems is that a strongly repulsive
bosonic system can be mapped to a noninteracting fermionic system ~\cite{tonks,bgir,ll,paredes,astrak-stg,Haller-supertonks}.
This gives rise to a strongly correlated phase known as the Tonks-Girardeau (TG)
gas. 
For spin-1/2 fermions in a one-dimensional trap, in the strongly repulsive limit, the spin-1/2 fermions
may form a ground state which is identical to that of noninteracting
fermions without spin \cite{girardeau, guan-supertonks}. 
High-precision control of such systems
has been proven feasible in a recent experiment which allows for preparing the
system in a state with a well-defined, small number of particles \cite{Jochim1}. In
particular, it has become possible to study the ground state and the dynamics of a
two-fermion system \cite{zuern}, for which the exact theoretical solution
is known in the full interaction parameter range \cite{busch,SowinskiTwoBosons,rontani}.

To describe systems with three or more fermions, different analytical
and numerical methods have been applied
\cite{deg,yang,brouzos,lindgren,bugnion,jho}, suggesting such
systems as a tool for studying ferromagnetism, and providing some insight into the
fermionized nature of the strongly repulsive system.
In this paper, we give a theoretical
description of few fermions in a one-dimensional harmonic trap based on an exact
diagonalization study. 
This allows us to go beyond the analytic solution of Ref. \cite{deg}, as we
cover the full energy spectrum in the full range of interaction strengths.
We focus on the quasi-degenerate regime where any small temperature or a
completely depolarizing channel may lead to an occupation of several states in
the spectrum. As a signature of this effect, we
calculate the occupation numbers of the harmonic oscillator levels, which are
found to significantly differ from the ground state expectation value. On the
other hand, as the true ground state is protected against mixing with other
states by permutation symmetry, such thermalized states require mechanisms to
break the degeneracy. While anharmonicities in the trap are found to fail, a
small
magnetic field gradient is shown to mix the degenerate states, giving rise to a
non-trivial spin dynamics.

\section{System} 
Our system consists of two-species fermions of mass $m$ confined in a one-dimensional trap with frequency $\omega$. The Hamiltonian has the form
\begin{equation} \label{Hamiltonian}
H=\sum_{i=1}^N\left[ -\frac{\hbar ^2}{2m}\frac{\partial ^2}{\partial x_i^2}
+\frac{m\omega^2}{2} x^2\right] +g_{1D}\sum_{i<j}\delta (x_i-x_j),
\end{equation}
where $g_{1D}$ is an effective interaction strength between two fermions of
different spins. In the following, we refer to the two species as a single species with an internal
(pseudo)spin-1/2 degree of freedom. We express all quantities in harmonic
oscillator units, i.e. $\hbar \omega$ for energy, $\sqrt{\hbar/m\omega}$ for
length, etc. For convenience we introduce the dimensionless interaction strength
$g=(m/\hbar^3\omega)^{1/2}\,g_{1D}$. 
Let us note that the interaction term in \eqref{Hamiltonian}
is non-zero only for states having a spatial wave function which is symmetric
under particle exchange. For fermions with the same spin the wave
function is always antisymmetric, and the interaction term will not contribute.
For two fermions with opposite spins, symmetric and antisymmetric wave functions
are possible and correspond to states with zero and finite interaction energy.

A convenient basis for studying the many-body problem is given by the
eigenstates $\phi_n(x)$ of the single-particle problem, simply being the
harmonic oscillator eigenfunctions corresponding to energies $\epsilon_n=n+1/2$.
The Hamiltonian \eqref{Hamiltonian} is then diagonalized in blocks with a fixed
total number of particles $N$, and fixed numbers $N_{\uparrow}$
($N_{\downarrow}$) of $\uparrow$ ($\downarrow$) fermions, defining the
$z$-component of the spin. We truncate the single-particle basis at a
sufficiently large level, $n_{\rm max}=20.$ 

\begin{figure}
\centering
\includegraphics{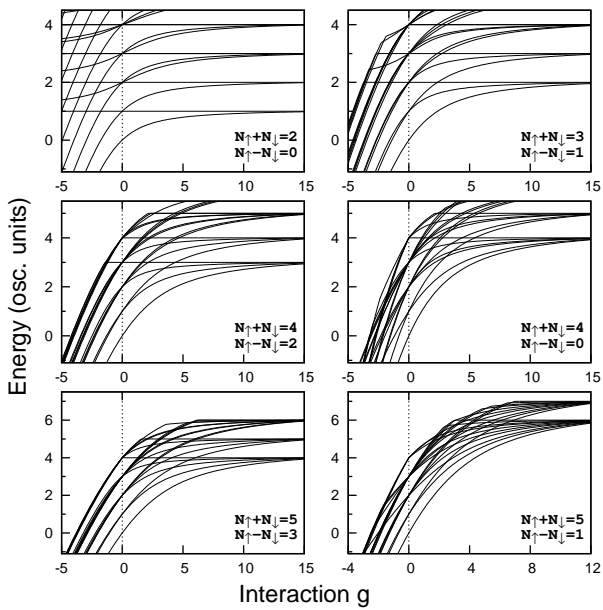}
\caption{We plot the energy $E-E_{\rm F}$ as a function of the dimensionless interaction
strength $g$ for different combinations of spin-up and spin-down particles. The
energy offset $E_{\rm F}$ is the Fermi energy of the non-interacting system. We
find ground state degeneracies in the limit of strong interactions. 
We plot first 20 eigenenergies of the Hamiltonian, in Hilbert spaces with
fixed $S_z=N_\uparrow-N_\downarrow$, but without fixing the total spin.}
\label{Fig1}
\end{figure}

Before turning to our numerical results, let us consider two limiting cases
which can be solved analytically. The first case is a non-interacting
system, $g=0$. The ground state is then obtained by simply filling the
Fermi sea, defining the Fermi energy $E_{\mathrm F}$. The second limiting case
is the Girardeau limit of infinitely strong repulsive interaction between the
two species, $g \rightarrow \infty$. Then, a Fermi-Fermi mapping \cite{ll},
allows one to treat the repulsive two-species fermions like non-interacting
one-species fermions. A spatial wave function for the ground state is then
obtained as a Slater determinant of the $N$ lowest levels. It can be rewritten
as \cite{girardeau}:
\begin{align}
\label{gir}
 \Psi \propto \left[\prod_{i=1}^{N}e^{-x_{i}^{2}/2}\right]
\prod_{1\le j<\ell\le N} (x_j-x_{\ell}).
\end{align}
This spatial wave function is fully antisymmetric and thus corresponds to a
fully symmetric spin configuration. It is an eigenfunction of both the
single-particle and the interaction part of the Hamiltonian and thus provides
an exact eigenfunction for any choice of $g$. In particular, as the wave
function vanishes whenever two particles are at the same position, it describes
a state with an energy which is independent from $g$, and which becomes the
ground state energy for $g \rightarrow \infty$.

It is possible to symmetrize Eq. (\ref{gir}) with respect to pairs of particles
of opposite spin just by including a factor ${\rm sgn}(x_{\ell}-x_k)$. However,
as our numerical results suggest, wave functions obtained in that way are
eigenfunctions of the Hamiltonian only for $g\rightarrow \infty$. More insight
is provided by an exact solution of the two-particle problem \cite{busch,SowinskiTwoBosons}
by rewriting the Hamiltonian into the relative motion $r=x_1-x_2$ and the center-of-mass motion $R=(x_1+x_2)/2$ coordinates. The relative motion of the two particles is then described by
\begin{align}
\label{Hrel}
 H_{\rm rel} = - \frac{ {\rm d}^2}{{\rm d}r^2} + \frac{1}{4} r^2 + g
\delta(r).
\end{align}
The relative motion part of the wave function \eqref{gir} is found to be the first
excited state of the Hamiltonian \eqref{Hrel} with energy $E-E_{\rm F}=1$ for
any $g$.
Its center-of-mass motion is in the ground state. In the limit $g\rightarrow
\infty$, the state \eqref{gir} becomes degenerate with the ground state which
smoothly evolves to the symmetric state $\Psi_0(r,R) \propto |r| e^{-r^2/4}
e^{-R^2}$, as we adiabatically increase $g$. Note that this wave function,
despite describing a state of zero interaction energy, is not an eigenstate of
Eq. (\ref{Hrel}) for any finite $g$.

We thus have seen that in the limit of infinitely strong interactions, the
two-particle problem has two degenerate ground states with opposite symmetry of
the spatial wave function. One solution is obtained from the other by
multiplying ${\rm sgn}(x_1-x_2)$. This operation turns the spatially
antisymmetric wave function \eqref{gir} into a spatially symmetric wave
function, and thus has to be accompanied with a corresponding change in the
symmetry of the spin wave function.

\begin{figure}
\centering
\includegraphics{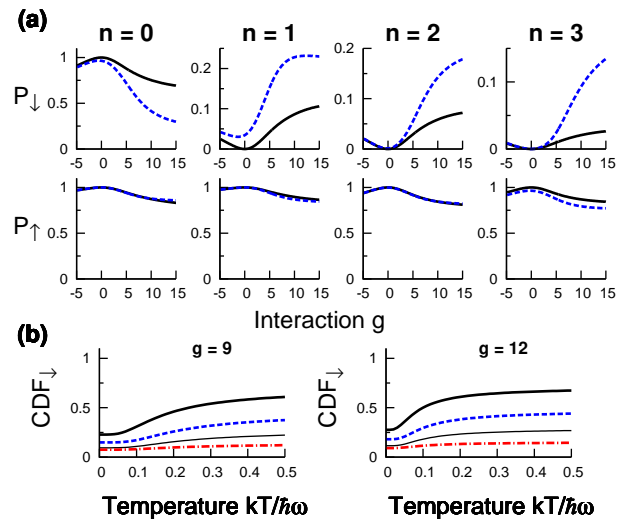}
\caption{(Color online) (a) The probabilities $P_{\downarrow} and P_{\uparrow}$ for finding $\uparrow$ and $\downarrow$ particles in
different orbitals ($n=0,1,2,3$) as a function of the dimensionless interaction strength $g$. We consider 
the effect of temperature using a Boltzman distribution: the
solid black and dashed blue lines denote the temperatures $kT/\hbar\omega = 0$ and $0.3$ respectively.
(b) Temperature dependence of the cumulative distribution function for $g=9$
(left) and $g=12$ (right). This function describes the probability of finding
the $\downarrow$ particle above the $n$th harmonic oscillator level, where $n=0$
is shown by the thick solid line, $n=1$ by the dashed line, $n=2$ by the thin solid
line, and $n=3$ by the dash-dotted line. 
In all plots $N=5$ and $N_\downarrow =1$. } 
\label{Fig2}
\end{figure}

The same mechanism can be applied for larger systems, $N>2$. Then, for every
pair of particles with opposite spin, it is possible to change the symmetry of
the spatial wavefunction in the state \eqref{gir} and thereby construct new
solutions in the Girardeau limit. This has been done in Ref. \cite{deg} and
leads to a degenerate ground-state manifold, where the number of
degenerate ground states $D$ is given by the number of distinct spin
configurations. It is counted by the distinct possibilities of dividing $N$
particles into two groups with $N_\uparrow$ and $N_\downarrow$ members; that is,
$D=\frac{N!}{N_{\downarrow}!N_{\uparrow}!}$. Note that the degeneracy of 
higher manifolds, corresponding to an excited center-of-mass motion, increases
since one also has to take into account excitations in the relative
motion.

\section{Role of degeneracies}
These degeneracies, although known before \cite{deg}, might play a crucial role in
understanding the few-body physics of strongly interacting fermions.
Our numerics focuses on the region between the two limiting cases, where, for
$N>2$, exact solutions are not known. In that region, the system makes use of
both the possibility of doubly occupying the lowest levels to reduce potential
energy, and occupying higher levels in order to reduce interactions. Energy
spectra as a function of interaction strength $g$ are plotted in Fig.
\ref{Fig1} for fixed $N_{\uparrow}$ and $N_{\downarrow}$. With this also the total
particle number and the $z$ component of spin, $S_z=N_\uparrow - N_\downarrow$, are
fixed, but not the total spin. We find different energy manifolds which become
degenerate in the limit $g \rightarrow \infty$. Each manifold corresponds to
different center-of-mass wave functions. The number of degenerate states in the
lowest manifold is given by $D$, the number of different spin configurations.
For any $S_z$, the highest energy state of the lowest manifold is described by
the fully antisymmetric wave function of Eq. (\ref{gir}). As explained above,
it is an exact solution with zero interaction energy for any $g$, and its energy
function is therefore simply a horizontal line. The degeneracy is lifted at any
finite $g$.

We next consider the population of the different single-particle levels.
In the quasidegenerate regime, the vanishing small energy gap does not protect
the ground state against mixing with other states from the manifold: If the
system's temperature is of the order of the gap, a description in terms
of thermal states becomes necessary. In Fig. \ref{Fig2}(a) we show, for
$N_{\uparrow}=4$ and  $N_{\downarrow}=1$, how temperature strongly
affects the occupation probabilities of the $\downarrow$ particle. Taking into
account the whole manifold of five quasi-degenerate states, we apply a Boltzmann
average at different temperatures. For $g\gtrsim 10$,  the probability of
finding the $\downarrow$ particle in the level $n=0$ is clearly reduced even by
a small temperature $k_{\rm B}T \ll \hbar\omega$. To understand this, we note
that at zero temperature, as shown by the thick solid line, only the true ground
state is occupied. For this state, the probability for the $\downarrow$ particle
to be in $n=0$ ranges between 1 at $g=0$ down to 0.8 for $g \gg 1$. In contrast,
the Girardeau state has an equal population of the five lowest level, such that
the probability of finding the $\downarrow$ particle in $n=0$ is given by 0.2.
This shows that the quasi-degeneracy of the ground state enhances the population
of the higher energy levels in the limit $g \rightarrow \infty$. Interestingly,
as shown by the second line of Fig. \ref{Fig2}(a), the occupation probabilities
for the $\uparrow$ particles are close to unity, independent of temperature
and interaction strength $g$, as a consequence of the Pauli principle.

From the experimental point of view, the probabilities shown in Fig.
\ref{Fig2}(a) cannot be measured directly. Instead, by tilting the trap
potential one can estimate the number of atoms above a certain harmonic oscillator level
by the counting atoms that leave the trap \cite{Jochim1}. In a series of measurements, 
this quantity is a counterpart of the cumulative distribution function
(CDF). This function describes the probability of a particle to occupy any
level above a certain cutoff level $n$. Measurements of CDFs for different $n$
would allow one to reproduce the probabilities of Fig. \ref{Fig2}(a). In Fig.
\ref{Fig2}(b), we plot the temperature dependence of the CDF of the $\downarrow$
particle in a system with $N_\downarrow=1$ and $N_\uparrow=4$  for two
different interaction strengths, $g=9$ and $g=12$. In both cases, the CDF of the
lowest energy levels ($n=0$ and $n=1$) is very sensible to small temperatures
and more than doubles in the plotted range $0\leq k_{\rm B}T \leq 0.5 \hbar
\omega$. In the case of $g=12$, this increase mostly takes place in the interval
$0\leq k_{\rm B}T \leq 0.15 \hbar \omega$, and the CDF saturates for larger
temperatures. This shows that temperature has become large compared to a
vanishingly small many-body gap, which exponentially decreases with $g$. Then,
the thermal regime transforms into the scenario where a completely depolarizing
channel simply favors the state with maximum entropy according to the Jaynes
principle.

\section{External symmetry breaking}
We now discuss the thermalization mechanisms that are able to bring the
system into a superposition of different states from the quasi-degenerate
manifold. On the basis of our ideal model, neither an adiabatic increase nor a
sudden quench of the interaction parameter would lead to occupation of more than
\textit{one} state in the degenerate manifold. A mixing with
other states is prohibited by the permutation-group symmetry, that is, the
conservation of total spin.

\begin{figure}
\centering
\includegraphics{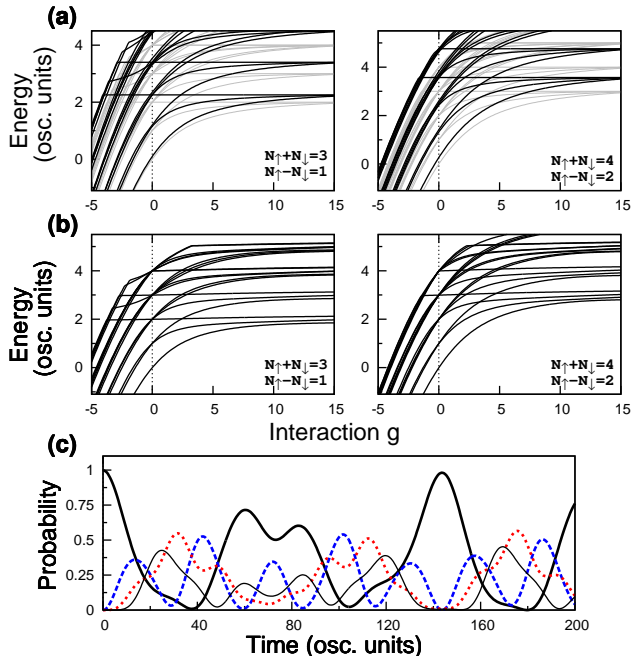}
\caption{(Color online) (a) Influence of a symmetric anharmonicity of the trap
to the spectrum of the Hamiltonian. Anharmonicity shifts eigenenergies of the
Hamiltonian (black lines) relative to the eigenenergies of the Hamiltonian with
the harmonic trap (gray lines). It does not lift the degeneracies at large $g$, and
provides no mixing between different states in each manifold. (b) Energies as a
function of $g$ in the presence of a Zeeman term. No degeneracies occur in the
strongly interacting limit.
(c) Zeeman breaking of the spin symmetry. For $N=4$ and $N_\uparrow=1$, the 
ground state at $g=12$ is time-evolved
after switching on a Zeeman term, $\delta = 0.05$. The probability for finding
the state in one of the four possible spin configurations is plotted. The thick
black, dashed blue, dotted red, and thin black lines correspond to the states
ordered by energy (increasing).  
}
\label{fig3}
\end{figure}

Also trap anharmonicities do not change this situation. As shown in Fig.
\ref{fig3}(a), they only shift the quasidegenerate energy manifolds
without lifting the degeneracy. In fact, the nature of the symmetry
dictates that thermalization mechanisms must simultaneously act on spin and
spatial degrees of freedom. This could be a spin-orbit coupling, spin-dependent
interaction like, for instance, $p$-wave interaction, or the existence of a
spatially dependent magnetic field. The latter option is easily implemented as
a Zeeman term, $H_{\rm Z} = \delta \sum_i x_i \sigma_i^z$, in the Hamiltonian.
Without loss of generality, we assume that the two internal states have
opposite magnetic moments along the $z$ direction and $\delta$ is a magnetic field
gradient. Such a term can be implemented in a controlled way in the experiment.

To study the effect of a Zeeman term in more detail, we first consider
two particles with opposite spins. It is easily seen that the
symmetric wave function $\ket{\rm S} \propto |r| e^{-r^2/4} e^{-R^2}$ (which
has to be multiplied by an antisymmetric spin wave function $\ket{\uparrow
\downarrow}-\ket{\downarrow \uparrow}$) has nonzero transition matrix elements
$\bra{\rm S} H_{\rm Z} \ket{\rm A}=2\sqrt{2/\pi}\delta$ with the antisymmetric
wave function  $\ket{\rm A} \propto r e^{-r^2/4} e^{-R^2}$ (which has to be
multiplied by a symmetric spin wave function $\ket{\uparrow
\downarrow}+\ket{\downarrow
\uparrow}$). This gives rise to a degeneracy splitting $\Delta$ which is linear
in the magnetic field gradient, $\Delta = 4\sqrt{2/\pi}\delta$, and to
mixed-symmetry states in the limit of large $g$.
The same effect is found for systems with three or four particles, where we have
taken into account the Zeeman term in our exact diagonalization study. The
situation is plotted in Fig. \ref{fig3}(b), clearly showing the lifted
degeneracy in the large $g$ limit. In particular, we find that for sufficiently
small Zeeman energy, $\delta \langle x \rangle \ll \hbar \omega$,
only states of the same energy manifold are mixed. Furthermore, since the Zeeman
term can be rewritten as a sum of operators acting only on pairs of particles, transition
matrix elements of states which differ by more than one unit of total spin are
zero. Accordingly, the matrix representation of the Zeeman term has a
tridiagonal structure. The mixed symmetry of the eigenstates in the presence of a Zeeman splitting is
illustrated by Fig. \ref{fig3}(c). For $N=4$ and $N_\uparrow=1$ we consider a
system which is prepared in the (maximum total spin) ground state of the
Hamiltonian for $g=12$ and $\delta=0$. Then we switch on the external magnetic
field gradient $\delta = 0.05$ and propagate the state for some time $t$. We
then measure the spin symmetry of the state (by projecting back into the
spin-conserving basis given by the eigenstates of $\delta=0$). We then plot the
probabilities of finding the system in one of the three different spin sectors
as a function of time. The time scale of the dynamics shown in Fig.
\ref{fig3}(c) can be controlled by the strength of the field gradient. This
could allow for studying the crossover from quantum time evolution to
thermalization.

\section{Conclusions}
In our study of one-dimensionally trapped spin-1/2 fermions we have focused on
the strongly repulsive regime, in which a ground-state degeneracy is exhibited.
This quasi-degenerate manifold allows one to study thermalization in a small
quantum-mechanical system. We have calculated correlation functions and
occupation numbers of the harmonic oscillator levels as signatures for 
distinguishing between pure states and thermal states
at finite or even infinite temperature. Since each eigenstate is protected from
mixing with other states by its symmetry with respect to the permutation group,
mechanisms for thermalization must in general be operators which
simultaneously act on spin and spatial degrees of freedom. We have shown that
the presence of an additional Zeeman term lifts the degeneracy, and may lead to
a time-dependent superposition of different states from the quasi-degenerate
manifold.

\textit{Note added in proof.}
Recently, we became aware of a related work by S. E. Gharashi
and D. Blume exploring the degenerate regime of strongly repulsive 1D fermions
\cite{blume}.

\section*{Acknowledgments}
We thank John Lapeyre for his helpful comments on the
manuscript, Selim Jochim for fruitful discussions, and Doerte Blume for
lending us their manuscript before publication. The work
has been supported by (Polish) National Science Center grants No.
DEC-2012/04/A/ST2/00090 (T.S.), DEC-2012/04/A/ST2/00088 (O.D.), Spanish MINCIN
project TOQATA (FIS2008-00784), ERC Advanced Grant QUAGATUA, EU IP SIQS, and EU
IP AQUTE. T.S. acknowledges support from the Foundation for Polish Science
(KOLUMB Programme; KOL/7/2012).

\bibliography{fermions}

\begin{thebibliography}{21}
\expandafter\ifx\csname natexlab\endcsname\relax\def\natexlab#1{#1}\fi
\expandafter\ifx\csname bibnamefont\endcsname\relax
  \def\bibnamefont#1{#1}\fi
\expandafter\ifx\csname bibfnamefont\endcsname\relax
  \def\bibfnamefont#1{#1}\fi
\expandafter\ifx\csname citenamefont\endcsname\relax
  \def\citenamefont#1{#1}\fi
\expandafter\ifx\csname url\endcsname\relax
  \def\url#1{\texttt{#1}}\fi
\expandafter\ifx\csname urlprefix\endcsname\relax\def\urlprefix{URL }\fi
\providecommand{\bibinfo}[2]{#2}
\providecommand{\eprint}[2][]{\url{#2}}

\bibitem[{\citenamefont{{M. Lewenstein} et~al.}(2012)\citenamefont{{M.
  Lewenstein}, {A. Sanpera}, and Ahufinger}}]{mlbook}
\bibinfo{author}{\bibnamefont{{M. Lewenstein}}},
  \bibinfo{author}{\bibnamefont{{A. Sanpera}}}, \bibnamefont{and}
  \bibinfo{author}{\bibfnamefont{V.}~\bibnamefont{Ahufinger}},
  \emph{\bibinfo{title}{Ultracold Atoms in Optical Lattices: Simulating Quantum
  Many-Body Systems}} (\bibinfo{publisher}{Oxford University Press (London)},
  \bibinfo{year}{2012}).

\bibitem[{\citenamefont{Tonks}(1936)}]{tonks}
\bibinfo{author}{\bibfnamefont{L.}~\bibnamefont{Tonks}},
  \bibinfo{journal}{Phys. Rev.} \textbf{\bibinfo{volume}{50}},
  \bibinfo{pages}{955} (\bibinfo{year}{1936}).

\bibitem[{\citenamefont{Girardeau}(1960)}]{bgir}
\bibinfo{author}{\bibfnamefont{M.}~\bibnamefont{Girardeau}},
  \bibinfo{journal}{J. Math. Phys.} \textbf{\bibinfo{volume}{1}},
  \bibinfo{pages}{516} (\bibinfo{year}{1960}).

\bibitem[{\citenamefont{Lieb and Liniger}(1963)}]{ll}
\bibinfo{author}{\bibfnamefont{E.}~\bibnamefont{Lieb}} \bibnamefont{and}
  \bibinfo{author}{\bibfnamefont{W.}~\bibnamefont{Liniger}},
  \bibinfo{journal}{Phys. Rev.} \textbf{\bibinfo{volume}{130}},
  \bibinfo{pages}{1605} (\bibinfo{year}{1963}).

\bibitem[{\citenamefont{Paredes et~al.}(2004)\citenamefont{Paredes, Widera,
  Murg, Mandel, Folling, Cirac, Shlyapnikov, H\"ansch, and Bloch}}]{paredes}
\bibinfo{author}{\bibfnamefont{B.}~\bibnamefont{Paredes}},
  \bibinfo{author}{\bibfnamefont{A.}~\bibnamefont{Widera}},
  \bibinfo{author}{\bibfnamefont{V.}~\bibnamefont{Murg}},
  \bibinfo{author}{\bibfnamefont{O.}~\bibnamefont{Mandel}},
  \bibinfo{author}{\bibfnamefont{S.}~\bibnamefont{Folling}},
  \bibinfo{author}{\bibfnamefont{I.}~\bibnamefont{Cirac}},
  \bibinfo{author}{\bibfnamefont{G.~V.} \bibnamefont{Shlyapnikov}},
  \bibinfo{author}{\bibfnamefont{T.~W.} \bibnamefont{H\"ansch}},
  \bibnamefont{and} \bibinfo{author}{\bibfnamefont{I.}~\bibnamefont{Bloch}},
  \bibinfo{journal}{Nature (London)} \textbf{\bibinfo{volume}{429}},
  \bibinfo{pages}{277} (\bibinfo{year}{2004}).

\bibitem[{\citenamefont{Astrakharchik et~al.}(2005)\citenamefont{Astrakharchik,
  Boronat, Casulleras, and Giorgini}}]{astrak-stg}
\bibinfo{author}{\bibfnamefont{G.~E.} \bibnamefont{Astrakharchik}},
  \bibinfo{author}{\bibfnamefont{J.}~\bibnamefont{Boronat}},
  \bibinfo{author}{\bibfnamefont{J.}~\bibnamefont{Casulleras}},
  \bibnamefont{and} \bibinfo{author}{\bibfnamefont{S.}~\bibnamefont{Giorgini}},
  \bibinfo{journal}{Phys. Rev. Lett.} \textbf{\bibinfo{volume}{95}},
  \bibinfo{pages}{190407} (\bibinfo{year}{2005}).

\bibitem[{\citenamefont{Haller et~al.}(2009)\citenamefont{Haller, Gustavsson,
  Mark, Danzl, Hart, Pupillo, and N\"agerl}}]{Haller-supertonks}
\bibinfo{author}{\bibfnamefont{E.}~\bibnamefont{Haller}},
  \bibinfo{author}{\bibfnamefont{M.}~\bibnamefont{Gustavsson}},
  \bibinfo{author}{\bibfnamefont{M.~J.} \bibnamefont{Mark}},
  \bibinfo{author}{\bibfnamefont{J.~G.} \bibnamefont{Danzl}},
  \bibinfo{author}{\bibfnamefont{R.}~\bibnamefont{Hart}},
  \bibinfo{author}{\bibfnamefont{G.}~\bibnamefont{Pupillo}}, \bibnamefont{and}
  \bibinfo{author}{\bibfnamefont{H.-C.} \bibnamefont{N\"agerl}},
  \bibinfo{journal}{Science} \textbf{\bibinfo{volume}{325}},
  \bibinfo{pages}{1224} (\bibinfo{year}{2009}).

\bibitem[{\citenamefont{Girardeau}(2010)}]{girardeau}
\bibinfo{author}{\bibfnamefont{M.~D.} \bibnamefont{Girardeau}},
  \bibinfo{journal}{Phys. Rev. A} \textbf{\bibinfo{volume}{82}},
  \bibinfo{pages}{011607(R)} (\bibinfo{year}{2010}).

\bibitem[{\citenamefont{Guan and Chen}(2010)}]{guan-supertonks}
\bibinfo{author}{\bibfnamefont{L.}~\bibnamefont{Guan}} \bibnamefont{and}
  \bibinfo{author}{\bibfnamefont{S.}~\bibnamefont{Chen}},
  \bibinfo{journal}{Phys. Rev. Lett.} \textbf{\bibinfo{volume}{105}},
  \bibinfo{pages}{175301} (\bibinfo{year}{2010}).

\bibitem[{\citenamefont{Serwane et~al.}(2011)\citenamefont{Serwane, Z\"urn,
  Lompe, Ottenstein, Wenz, and Jochim}}]{Jochim1}
\bibinfo{author}{\bibfnamefont{F.}~\bibnamefont{Serwane}},
  \bibinfo{author}{\bibfnamefont{G.}~\bibnamefont{Z\"urn}},
  \bibinfo{author}{\bibfnamefont{T.}~\bibnamefont{Lompe}},
  \bibinfo{author}{\bibfnamefont{T.~B.} \bibnamefont{Ottenstein}},
  \bibinfo{author}{\bibfnamefont{A.~N.} \bibnamefont{Wenz}}, \bibnamefont{and}
  \bibinfo{author}{\bibfnamefont{S.}~\bibnamefont{Jochim}},
  \bibinfo{journal}{Science} \textbf{\bibinfo{volume}{332}},
  \bibinfo{pages}{6027} (\bibinfo{year}{2011}).

\bibitem[{\citenamefont{Z\"urn et~al.}(2012)\citenamefont{Z\"urn, Serwane,
  Lompe, Wenz, Ries, Bohn, and Jochim}}]{zuern}
\bibinfo{author}{\bibfnamefont{G.}~\bibnamefont{Z\"urn}},
  \bibinfo{author}{\bibfnamefont{F.}~\bibnamefont{Serwane}},
  \bibinfo{author}{\bibfnamefont{T.}~\bibnamefont{Lompe}},
  \bibinfo{author}{\bibfnamefont{A.~N.} \bibnamefont{Wenz}},
  \bibinfo{author}{\bibfnamefont{M.~G.} \bibnamefont{Ries}},
  \bibinfo{author}{\bibfnamefont{J.~E.} \bibnamefont{Bohn}}, \bibnamefont{and}
  \bibinfo{author}{\bibfnamefont{S.}~\bibnamefont{Jochim}},
  \bibinfo{journal}{Phys. Rev. Lett.} \textbf{\bibinfo{volume}{108}},
  \bibinfo{pages}{075303} (\bibinfo{year}{2012}).

\bibitem[{\citenamefont{Busch et~al.}(1998)\citenamefont{Busch, Englert,
  Rz\k{a}\.zewski, and Wilkens}}]{busch}
\bibinfo{author}{\bibfnamefont{T.}~\bibnamefont{Busch}},
  \bibinfo{author}{\bibfnamefont{B.-G.} \bibnamefont{Englert}},
  \bibinfo{author}{\bibfnamefont{K.}~\bibnamefont{Rz\k{a}\.zewski}},
  \bibnamefont{and} \bibinfo{author}{\bibfnamefont{M.}~\bibnamefont{Wilkens}},
  \bibinfo{journal}{Foundations of Physics} \textbf{\bibinfo{volume}{28}},
  \bibinfo{pages}{549} (\bibinfo{year}{1998}).

\bibitem[{\citenamefont{Sowi\ifmmode~\acute{n}\else \'{n}\fi{}ski
  et~al.}(2010)\citenamefont{Sowi\ifmmode~\acute{n}\else \'{n}\fi{}ski,
  Brewczyk, Gajda, and Rz\k{a}\ifmmode~\dot{z}\else
  \.{z}\fi{}ewski}}]{SowinskiTwoBosons}
\bibinfo{author}{\bibfnamefont{T.}~\bibnamefont{Sowi\ifmmode~\acute{n}\else
  \'{n}\fi{}ski}}, \bibinfo{author}{\bibfnamefont{M.}~\bibnamefont{Brewczyk}},
  \bibinfo{author}{\bibfnamefont{M.}~\bibnamefont{Gajda}}, \bibnamefont{and}
  \bibinfo{author}{\bibfnamefont{K.}~\bibnamefont{Rz\k{a}\ifmmode~\dot{z}\else
  \.{z}\fi{}ewski}}, \bibinfo{journal}{Phys. Rev. A}
  \textbf{\bibinfo{volume}{82}}, \bibinfo{pages}{053631}
  (\bibinfo{year}{2010}).

\bibitem[{\citenamefont{Rontani}(2012)}]{rontani}
\bibinfo{author}{\bibfnamefont{M.}~\bibnamefont{Rontani}},
  \bibinfo{journal}{Phys. Rev. Lett.} \textbf{\bibinfo{volume}{108}},
  \bibinfo{pages}{115302} (\bibinfo{year}{2012}).

\bibitem[{\citenamefont{Guan et~al.}(2009)\citenamefont{Guan, Chen, Wang, and
  Ma}}]{deg}
\bibinfo{author}{\bibfnamefont{L.}~\bibnamefont{Guan}},
  \bibinfo{author}{\bibfnamefont{S.}~\bibnamefont{Chen}},
  \bibinfo{author}{\bibfnamefont{Y.}~\bibnamefont{Wang}}, \bibnamefont{and}
  \bibinfo{author}{\bibfnamefont{Z.-Q.} \bibnamefont{Ma}},
  \bibinfo{journal}{Phys. Rev. Lett.} \textbf{\bibinfo{volume}{102}},
  \bibinfo{pages}{160402} (\bibinfo{year}{2009}).

\bibitem[{\citenamefont{Yang}(2009)}]{yang}
\bibinfo{author}{\bibfnamefont{C.~N.} \bibnamefont{Yang}},
  \bibinfo{journal}{Chinese Physics Letters} \textbf{\bibinfo{volume}{26}},
  \bibinfo{pages}{120504} (\bibinfo{year}{2009}).

\bibitem[{\citenamefont{Brouzos and Schmelcher}(2013)}]{brouzos}
\bibinfo{author}{\bibfnamefont{I.}~\bibnamefont{Brouzos}} \bibnamefont{and}
  \bibinfo{author}{\bibfnamefont{P.}~\bibnamefont{Schmelcher}},
  \bibinfo{journal}{Phys. Rev. A} \textbf{\bibinfo{volume}{87}},
  \bibinfo{pages}{023605} (\bibinfo{year}{2013}).

\bibitem[{\citenamefont{Lindgren et~al.}(2013)\citenamefont{Lindgren, Rotureau,
  Forss\'en, Volosniev, and Zinner}}]{lindgren}
\bibinfo{author}{\bibfnamefont{E.~J.} \bibnamefont{Lindgren}},
  \bibinfo{author}{\bibfnamefont{J.}~\bibnamefont{Rotureau}},
  \bibinfo{author}{\bibfnamefont{C.}~\bibnamefont{Forss\'en}},
  \bibinfo{author}{\bibfnamefont{A.~G.} \bibnamefont{Volosniev}},
  \bibnamefont{and} \bibinfo{author}{\bibfnamefont{N.~T.}
  \bibnamefont{Zinner}}, \bibinfo{journal}{arXiv:1304.2992}
  (\bibinfo{year}{2013}).

\bibitem[{\citenamefont{Bugnion and Conduit}(2013)}]{bugnion}
\bibinfo{author}{\bibfnamefont{P.~O.} \bibnamefont{Bugnion}} \bibnamefont{and}
  \bibinfo{author}{\bibfnamefont{G.~J.} \bibnamefont{Conduit}},
  \bibinfo{journal}{arXiv:1304.3299}  (\bibinfo{year}{2013}).

\bibitem[{\citenamefont{Cui and Ho}(2013)}]{jho}
\bibinfo{author}{\bibfnamefont{X.}~\bibnamefont{Cui}} \bibnamefont{and}
  \bibinfo{author}{\bibfnamefont{T.-L.} \bibnamefont{Ho}},
  \bibinfo{journal}{arXiv:1305.6361}  (\bibinfo{year}{2013}).

\bibitem[{\citenamefont{Gharashi and Blume}(2013)}]{blume}
\bibinfo{author}{\bibfnamefont{S.~E.} \bibnamefont{Gharashi}} \bibnamefont{and}
  \bibinfo{author}{\bibfnamefont{D.}~\bibnamefont{Blume}},
  \bibinfo{journal}{Phys. Rev. Lett.} \textbf{\bibinfo{volume}{111}},
  \bibinfo{pages}{045302} (\bibinfo{year}{2013}).

\end{thebibliography}

\end{document}